\documentclass[floatfix,aps,prd,showpacs,amsmath,amssymb,nofootinbib,
preprintnumbers]{revtex4}

\begin{document}

\title{Disappearing cosmological constant in $f(R)$ gravity}

\author{Alexei A. Starobinsky}
\email{alstar@landau.ac.ru}

\affiliation{Landau Institute for Theoretical Physics, Russian Academy 
of Sciences, Moscow 119334, Russia, and \\
Yukawa Institute for Theoretical Physics, Kyoto University, Kyoto 
606-8502, Japan}

\date{June 14, 2007}

\begin{abstract}

For higher-derivative $f(R)$ gravity where $R$ is the Ricci scalar,
a class of models is proposed which produce viable cosmology different
from the LambdaCDM one at recent times and satisfy cosmological, Solar 
system and laboratory tests. These models have both flat and de Sitter
space-times as particular solutions in the absence of matter. Thus, 
a cosmological constant is zero in flat space-time, but appears 
effectively in a curved one for sufficiently large $R$. A 'smoking gun' 
for these models would be small discrepancy in values of the slope of 
the primordial perturbation power spectrum determined from galaxy 
surveys and
CMB fluctuations. On the other hand, a new problem for dark energy 
models based on $f(R)$ gravity is pointed which is
connected with possible overproduction of new massive scalar particles
(scalarons) arising in this theory in the very early Universe. 

\end{abstract}


\maketitle

\section{Introduction}

Continuing investigation of dark energy (DE) properties in the Universe
(see the recent review \cite{SS06} for the definitions of what is usually
called the effective DE energy density $\rho_{DE}$ and pressure $p_{DE}$ 
from the observational point of view) has shown that its properties are 
very close to those of an exact cosmological constant $\Lambda$ that has 
$\rho_{\Lambda}= - p_{\Lambda}= \Lambda/8\pi G=const>0$.\footnote {The
sign conventions are: the metric signature $(+---)$, the curvature tensor
$R^{\sigma}_{~\mu\rho\nu}=\partial_{\nu}\Gamma^{\sigma}_{\mu\rho} - ...,~
R_{\mu\nu}=R^{\sigma}_{~\mu\sigma\nu}$, so that the Ricci scalar $R\equiv
R^{\mu}_{\mu}>0$ for the de Sitter space-time and the matter-dominated 
cosmological epoch; $c=\hbar=1$ is assumed throughout the paper.} In
particular, if $w_{DE}\equiv p_{DE}/\rho_{DE}$ is assumed to be
constant, then $|w_{DE}+1|<0.1$ ($1\sigma$ error bars) or even smaller,
see \cite{recent} for the analysis of the most recent observational
data using different techniques.\footnote{Note that the assumption 
$w_{DE}=const<0$ and not equal to $-1,-2/3,-1/3$ is not very natural
in the presence of non-relativistic matter since it requires DE
models with rather specific potentials, see e.g. \cite{SS00}.} 
However, for more generic DE models
with $w_{DE}\not= const$, the same analysis does not exclude varying
$w_{DE}$ including even temporal phantom behaviour of DE ($w_{DE}<-1$)
at recent redshifts $z<0.3$. The latter behaviour (in other words,
breaking of the weak energy condition for DE), if confirmed by future,
more exact data, may not be explained in the scope of physical DE
models according to the terminology of \cite{SS06} (e.g., the
quintessence ones) and requires some kind of geometrical DE (otherwise 
dubbed modified gravity), see also recent reviews \cite{CST06}.   
 
Among geometrical DE models one of the most simplest ones is the
$f(R)$ class of gravity models with the Lagrangian density 
\begin{equation}
L= \frac {f(R)}{16\pi G} + L_m~,
\label{Lagr}
\end{equation}
where $L_m$ describes all non-gravitational kinds of matter including 
non-relativistic (cold) dark matter and the metric variation is 
assumed.\footnote{The Palatini variation of (\ref{Lagr}) leads to 
completely different equations of motion, even the number of degrees
of freedom (i.e. particle content) is not the same. Thus, two
models with the same Lagrangian density (\ref{Lagr}) but different 
ways of variation should be studied and compared with data as two 
separate models; their similarity is illusory.} $f(R)$ may be an 
arbitrary function subjected, however, to some stability conditions 
discussed below. For $f(R) = R - 2\Lambda$, it 
reduces to the Einstein gravity with a cosmological constant. Thus, it 
contains the standard cosmological $\Lambda$CDM model as a particular 
case. However for $f''(R)\not= 0$, in addition to the massless spin-2 
graviton, this class of models contains a scalar particle, dubbed 
scalaron in \cite{S80}, which rest-mass is $M^2(R)=(3f''(R))^{-1}$ in 
the WKB-regime $|M^2|\gg R^2,~R_{\mu\nu}R^{\mu\nu}$ (here and below 
prime means differentiation with respect to an argument). It is neither 
a tachyon, not a ghost for $f''(R)>0$ in this regime. Finally, graviton
is not a ghost if $f'(R)>0$. All these properties can be easily 
obtained either directly, or (in the absence of $L_m$) using conformal
equivalence of equations of motions for this class of models to those 
of the Einstein gravity interacting with a minimally coupled scalar
field $\phi$ with some potential $V(\phi)$ which form is uniquely
determined by $f(R)$ in all points where $f'(R)\not= 0$ \cite{W84}. 
However, I shall not use this conformal equivalence below, in 
particular, because it may be misleading in the presence of other
kinds of matter. 

In this paper Eq. (\ref{Lagr}) is considered as a purely 
phenomenological effective Lagrangian density describing geometrical DE. 
However, in principle, it may arise either due to quantum fluctuations 
of all fields including gravity (as was supposed in early papers like 
\cite{S80}), or as a result of reduction from higher dimensions to 4D
in some variant of modern string/M-theory (see e.g. \cite{NO03} in 
this respect).   

First papers on cosmological models in $f(R)$ gravity appeared
already in 1969-1970 \cite{RR69}. Then, among other results, this
class of models with $f(R)=R+R^2/6M^2$ plus some small non-local
terms (which are crucial for reheating after inflation) was used to 
construct the first internally self-consistent cosmological model
possessing a (quasi-)de Sitter (latter dubbed inflationary) stage in 
the early Universe with slow-roll decay, a graceful exit to the
subsequent radiation-dominated Friedmann-Robertson-Walker (FRW) stage 
(through an intermediate matter-dominated one) and sufficiently
effective reheating in the regime of narrow parametric resonance
\cite{S80} (see \cite{S82} for more details). It is for this model
that calculations of scalar (adiabatic) perturbations generated
during inflation were first done \cite{MC81}. Moreover, with all 
recent observational data taken into account, it still remains among 
viable cosmological models: as follows from the final results for 
both scalar and tensor perturbations \cite{S83} (see also 
\cite{KMP87,HN01,FTBM06}), the
model predictions for the slope of the primordial spectrum of scalar
perturbations $n_s$ and the tensor/scalar ratio $r$ are $n_s-1=-2N^{-1} 
= -0.04~(N/50)^{-1},~r=12N^{-2}= 0.0048~(N/50)^{-2}$ where $N$ is the 
number of e-folds between the first Hubble radius crossing of the 
present inverse comoving scale $0.05$ Mpc$^{-1}$ and the end of 
inflation -- well in agreement with the present data (of course, in 
the case of $r$ we have an upper limit only). The only free parameter 
of this model -- the scalaron rest-mass $M$ -- is determined from the 
normalization of the primordial scalar spectrum. If we
take it from the best fit to the combined WMAP3-SDSS measurements 
\cite{T06}, then $ M=2.8\times 10^{-6}~(N/50)^{-1}M_{Pl}$ where
$M_{Pl}=1/\sqrt{G}$ (by the way, this lies inside the range 
conjectured in \cite{S83}).

Due to remarkable {\em qualitative} similarity between the 
present DE and primordial DE that supported inflation in the early 
Universe, all inflationary models may be applied to the description 
of the present DE, too, after changing numerical values of their 
microscopical parameters only.\footnote{This does not mean that the 
present and primordial DE should necessarily be the same kind of matter, 
like in the so called quintessential inflationary models. I am only 
speaking about possibility to use the same kind of theoretical models 
in both cases.} Actually, it had been done already for practically all 
models (sometimes in the inverse historical order). The same 
occurred to models based on $f(R)$ gravity beginning from \cite{C02}, 
and then it was proposed in \cite{CCT03} to use $f(R)$ models with
$f(R)$ diverging (or finite but non-analytic) at $R\to 0$ for 
description of the present DE. However, after much agitation on this 
particular class of models, it was proven that they are either 
non-viable, or practically indistinguishable from the standard 
$\Lambda$CDM model, see \cite{APT07,FTBM06} for such rather pessimistic 
conclusions as well as for extensive lists of publications on this 
topic. 

This does not completely close the way to construct a viable DE model 
in $f(R)$ gravity observationally distinguishable from the $\Lambda$CDM 
model, but suggests to abandon the hypothesis of divergence of $f(R)$
at $R=0$, as well at any other value of $R$, and to return to the 
natural assumption that $f(R)$ is regular
in this point. Moreover, an interesting and intriguing possibility  
is $f(0)=0$ while $f\to R-2\Lambda$ for $R\gg \Lambda$. This behaviour
corresponds to an effective cosmological constant existing in a
sufficiently curved space-time but 'disappearing' in the flat one --
that explains the title of the paper. In other words, in such a model
the observed DE (close to the cosmological constant for sufficiently
large $R$) is a purely curvature induced effect. It is 
totally unrelated to quantum vacuum energy in flat space-time 
that should be zero due to some other symmetry.\footnote{So, this 
model easily realizes  a possibility of 'degravitation of the 
cosmological constant' in the scalar sector which was recently 
proposed and tried to be achieved in the tensor sector in \cite{DHK07}
but faced with very serious technical problems due to non-locality of 
equations.} However, the price to pay is that flat space-time becomes
unstable with a characteristic time of the order of the present Universe
age. 

Of course, it is much more difficult to construct a viable DE
model in $f(R)$ gravity as compared to general scalar-tensor gravity
since the former contains only one arbitrary function while the latter
has two functions and provides much place for viable DE models, see  
e.g. \cite{BEPS00} for reconstruction of such models from different 
kinds of observational data and \cite{GPRS06} for models which admit 
recent phantom behaviour of DE. Another source of problems is that it 
is rather non-trivial 
to satisfy laboratory and Solar system tests in $f(R)$ gravity since 
it formally represents the limiting case $\omega=0$ of scalar-tensor
gravity where $\omega$ is the Brans-Dicke parameter \cite{TT83}, 
though sometimes this limit should be taken carefully (see  
\cite{CTE06} for recent reconsideration). However, this may be
considered even as an advantage of DE $f(R)$ models since it makes 
easier to falsify them. So, in the next section a trial 3-parametric 
form of $f(R)$ is introduced which realizes the 'disappearing 
cosmological constant' possibility and behaviour of its FRW solutions
is investigated. In Sec. 3 laboratory and Solar system tests, as well
as dynamics of small perturbations are considered.
Sec. 4 contains conclusions and discussion of problems and further 
tests of this model.

\section{Geometrical dark energy model and behaviour of its FRW 
solutions}

Field equations following from (\ref{Lagr}) can be written in the
following Einsteinian form (though gravity itself is not
the Einstein one):
\begin{equation}
R_{\mu}^{\nu}-\frac{1}{2}\delta_{\mu}^{\nu}R = - 8\pi G \left(
T_{\mu (m)}^{\nu} + T_{\mu (DE)}^{\nu}\right)
\label{Eeq}
\end{equation}
where 
\begin{equation}
8\pi GT_{\mu (DE)}^{\nu}\equiv F'(R) R_{\mu}^{\nu}- \frac{1}{2}
F(R)\delta_{\mu}^{\nu} +\left(\nabla_{\mu}\nabla^{\nu}- 
\delta_{\mu}^{\nu}\nabla_{\rho}\nabla^{\rho}\right) F'(R)~,
~~~F(R)\equiv f(R)-R
\label{TDE}
\end{equation}
and $T_{\mu (m)}^{\nu}$ follows from variation of $L_m$ and satisfies
the generalized conservation law $T^{\nu}_{\mu ;\nu (m)}=0$
separately (since the left-hand side of Eq. (\ref{Eeq}) and the
right-hand side of Eq. (\ref{TDE}) satisfy this condition, too).
There exists a subtlety in this representation that is discussed below.
The trace of Eq. (\ref{Eeq}) reads
\begin{equation}
3\nabla_{\mu}\nabla^{\mu}f' - Rf'+2f=8\pi GT_m~.
\label{trace}
\end{equation}
Constant curvature solutions (de Sitter ones for $R>0$) are roots
of the algebraic equation $Rf'=2f$.

Let us take $f(R)$ in the following 3-parametric form:
\begin{equation}
f(R)=R+\lambda R_0\left(\left(1+\frac{R^2}{R_0^2}\right)^{-n}-1\right)
\label{f}
\end{equation}
with $n,\lambda>0$ and $R_0$ of the order of the presently observed
effective cosmological constant.
Then $f(0)=0$ (the cosmological constant 'disappears' in flat space-time) 
and $R_{\mu}^{\nu}=0$ is always a solution of Eq. (\ref{Eeq}) in the
absence of matter, but $f''(0)$ is negative -- flat space-time is 
unstable. For $|R|\gg R_0$, $f(R)=R-2\Lambda(\infty)$ where the 
high-curvature value of the effective cosmological constant is
$\Lambda(\infty)=\lambda R_0/2$. The equation for de Sitter 
solutions having $R=const=R_1=x_1R_0,~x_1>0$ can be written in the form
\begin{equation}
\lambda=\frac{x_1(1+x_1^2)^{n+1}}{2\left((1+x_1^2)^{n+1}-1-
(n+1)x_1^2\right)}~.
\label{dS}
\end{equation}
Below, by $x_1$ I will mean the {\em maximal} root of Eq. (\ref{dS}). 
So, instead of specifying $\lambda$, one may take any value of $x_1$
and then determine the corresponding value of $\lambda$. It follows
from the structure of Eq. (\ref{dS}) that $x_1<2\lambda$. Thus,
the effective cosmological constant at the de Sitter solution
$\Lambda(R_1)=R_1/4<\Lambda(\infty)$. On the other hand, $x_1\to 
2\lambda$ in both limiting cases $x_1$ fixed, $n\gg 1$ and 
$x_1\gg 1$, $n$ fixed. In these cases the Universe evolution 
becomes indistinguishable from that in the $\Lambda$CDM model.

Let us now consider the stability conditions
\begin{equation}
f'(R)>0~,~~f''(R)>0~,~~R\ge R_1~.
\label{stab}
\end{equation}
Note that they are imposed not in the whole space of solutions but only
on a trajectory of the evolution of our Universe from very large and 
positive $R$ in the past to $R=R_1$ in the infinite future. The 
quantum meaning of these conditions (graviton is not a ghost, scalaron
is not a tachyon) has already been mentioned in Sec. 1. However, 
violation of these conditions in the course of purely classical 
evolution is undesirable, too. If $f'(R)=0$ for some finite $R>R_1$,
a universe generically becomes strongly anisotropic and inhomogeneous
at some finite moment of time \cite{H73,GS79} (the same happens to the
Einstein gravity + a non-minimally coupled scalar field, 
preventing $G_{eff}$ from changing sign \cite{S81}). In the point 
where $f''(R)=0$, some weak singularity occurs which will be
considered elsewhere. In terms of the conformal equivalence mentioned
above, $dR/d\phi$ diverges at this point.

It can be shown that it is sufficient to satisfy the conditions 
(\ref{stab}) for $R=R_1$ and then they will be valid over the whole 
interval $[R_1,\infty)$. Correspondingly, this gives two necessary 
conditions for parameters of the form (\ref{f}):
\begin{equation}
(1+x_1^2)^{n+1}>1+(2n+1)x_1^2~,~~~x_1^2>1/(2n+1)~.
\label{ineq1}
\end{equation} 
 
To these inequalities, the condition of the stability of the future
de Sitter stage has to be added. It follows from variation of
Eq. (\ref{trace}) and reads (since the condition $f''(R_1)>0$ is 
already assumed to be satisfied) \cite{F05}:
\begin{equation}
f'(R_1)>R_1f''(R_1)~.
\label{stabds}
\end{equation}
This condition is stronger than the first of inequalities 
(\ref{stab}) at $R=R_1$, so it substitutes it. For our model it 
produces the requirement:
\begin{equation}
(1+x_1^2)^{n+2}> 1+(n+2)x_1^2+(n+1)(2n+1)x_1^4~.
\label{ineq2}
\end{equation}
It can be proven that it implies the second of inequalities 
(\ref{ineq1}), too. So, it is sufficient to check this inequality 
only. In particular, for $n=1$ it reads $x_1>\sqrt 3$ that leads to
$\lambda > 8/3\sqrt 3$. In addition, the value of $x_1$ saturating
(\ref{ineq2}) is also the point where $\lambda(x_1)$ in Eq. (\ref{dS})
reaches minimum.
  
Now turn to the model evolution at the matter dominated stage in 
the regime $R\gg R_0$. By construction, the model (\ref{f}) satisfies
the conditions 
\begin{equation}
|F|\ll R,~~|F'(R)|\ll 1,~~R|F''(R)|\ll 1
\label{iter}
\end{equation}
for $R\gg R_0$. Due to this, it is possible to solve Eq. (\ref{trace})
iteratively in this regime and corrections to the standard Einsteinian
(actually, even Newtonian) behaviour $R=R^{(0)}=8\pi GT_{m}\propto
a^{-3}$ appear to be small ($a(t)$ is a FRW scale factor).  
Thus, the model does not possess the Dolgov-Kawasaki instability 
\cite{DK03}. These corrections are of two types -- matter induced ones 
and free scalaron oscillations. The former ones follow from direct
iteration of Eq. (\ref{trace}):
\begin{equation}
R=R^{(0)}+\delta R_{ind}+\delta R_{osc}~,~~ 
\delta R_{ind}=\left(RF'(R)-2F(R)-3\nabla_{\mu}\nabla^{\mu}F'(R)
\right)_{R=R^{(0)}} 
\label{ind} 
\end{equation}
where $\nabla_{\mu}$ is taken with respect to the unperturbed metric, 
too. For $R\gg R_0$, $\delta R_{ind}\approx const=-2F(\infty)=
2\lambda R_0 = 4\Lambda(\infty)$ and $8\pi GT_{\mu (DE)}^{\nu}\approx
\Lambda(\infty)\delta_{\mu}^{\nu}$. Thus, DE behaves as a positive 
cosmological constant at redsifts $z\gg 1$.

Let us now return to the subtle point in the definition of the DE 
energy-momentum tensor (\ref{TDE}) pointed above. It consists in the 
following: what is the constant $G$ in it? For the model involved, $G$ 
coincides with $G_{eff}(\infty)$ -- the value of the Newtonian
gravitational constant measured in a Cavendish-type
experiment made in an environment with space-time curvature $R\gg R_0$.
In the next section, we will discuss under what restriction on $n$
this is achieved in laboratory experiments already made. Moreover,
it coincides with $G_{eff}$ at the radiation-dominated stage in the
Universe during the period of the Big Bang nucleosynthesis (BBN) 
$t\sim (1-100)$ s.\footnote{Note that at $t\sim 100$ s, after
antimatter -- positrons -- annihilation, $T_m$ is 
$\sim 10^{-3}$ g cm$^{-3}$ only, similar to conditions on the Earth.} 
So, BBN predictions remain unchanged in this DE model, 
too. Thus, such a choice of $G$ in the left-hand side of (\ref{TDE})
is very natural for this model. Taking another constant, say
$G(df/dR)_{now}^{-1}$ where the subscript 'now' means the present 
moment, results in adding a 'tracking' component to DE which is 
proportional to the Einstein tensor and may have any sign. That is why
with our definition (\ref{TDE}), there is no effect of $\rho_{DE}$ 
becoming negative at sufficiently large redshifts (with $w_{DE}$ 
diverging at the moment when $\rho_{DE}=0$) like that proposed in the 
recent paper \cite{AT07}. 
   
More unexpected is the behaviour of the second small correction 
$\delta R_{osc}$. It satisfies the equation 
\begin{equation}
\frac{3}{a^3}\frac{d}{dt}\left(a^3\frac{d}{dt}\left(F''(R^{(0)})\,
\delta R_{osc}\right)\right)+\delta R_{osc}=0~.
\label{eq-osc}
\end{equation}
For $R\gg R_0$, the WKB-approximation may be used. Then the solution
is
\begin{equation}
\delta R_{osc}=Ca^{-3/2}(F''(R^{(0)}))^{-3/4}\sin \left(\int 
\frac{dt}{\sqrt{3F''(R^{(0)})}}\right) 
\label{osc}
\end{equation}
with $C=const$. The integral here is just $\int M(R^{(0)})\, dt$ where 
$M$ is the scalaron mass introduced in Sec. 1. At the matter-dominated 
regime:
\begin{equation}
a \propto t^{2/3},~~R^{(0)}=\frac{4}{3t^2}~,~~F''\propto t^{4n+4}~,
~~M(R)\propto t^{-2n-2}~,
\label{mat-back}
\end{equation}
\begin{equation}
\delta R_{osc}\propto t^{-3n-4}\sin\left(const\cdot t^{-2n-1}\right)~,~~
\frac{\delta a}{a}\propto t^n\sin\left(const\cdot t^{-2n-1}\right)~.
\label{asymp}
\end{equation}
Similar behaviour continues during the radiation-dominated stage:
\begin{equation}
R^{(0)}\propto t^{-3/2}~,~~M(R)\propto t^{-3(n+1)/2}~,~~
\delta R_{osc}\propto t^{-\frac{9n}{4}-3}\sin\left(const\cdot
t^{-(3n+1)/2}\right)~.
\label{rad-as}
\end{equation} 
Thus, though oscillations of the scale factor remain small as $t\to 0$,
oscillations of $R$ grow to the past and finally violate the assumption
$|\delta R_{osc}|\ll R^{(0)}$. Note that the energy of scalaron
oscillations at that moment $\rho_{osc}\sim (\delta R_{osc})^2/GM^2(R)$
is still much less than $T_m$ since $R\ll M^2$ for $R\gg R_0$. 
Investigation of further FRW evolution of the model to the past is 
blocked by the stability problem: $R$ can become less than $R_1$ and
even negative during oscillations, so the conditions (\ref{stab}) are 
violated. For $n>1$, it is possible to choose $x_1$ or $\lambda$ in
such a way that $f'(R)>0$ for all real $R$. But $f''(R)$ always
becomes zero at $R=\pm R_2$ where $R_2=R_0/\sqrt{2n+1}<R_1$ and 
negative for $|R|<R_2$, and it is not possible to avoid this property
without abandoning the assumption of the cosmological constant 
disappearance in flat space-time $f(0)=0$.

Therefore, first, to avoid $M(R)$ becoming too large, say larger than 
$M_{Pl}$, the function $f(R)$ in (\ref{f}) has to be modified at 
$R\to\infty$. The simplest way is to add the term $R^2/6M^2$ to
(\ref{f}) where the value of $M$ (which will be the limiting value 
of $M(R)$ for $R\to \infty$) may be taken just that which is needed 
for the $R+R^2$ inflationary model mentioned in Sec. 1. This term will
be negligible for the present DE.

Second, a new serious problem for DE models in $f(R)$ gravity appears 
which has not been considered before: to avoid destroying radiation-
and matter-dominated FRW stages, some mechanism in the early Universe
should work to prohibit overproduction of scalarons. In the $R+R^2$ 
inflationary model such mechanism does exist -- gravitational creation 
of non-conformally-invariant particles and antiparticles (though not 
gravitons!) by oscillations of $R$. However, in the model (\ref{f}) it is 
not possible to use it in full until the question how to evolve through 
the point where $f''(R)=0$ is solved. The only way to avoid this
problem at all is to assume that $|\delta R_{osc}|< 8\pi GT_m$ just 
from the very beginning of early evolution of the Universe (or at least
from the moment when the effective Lagrangian density (\ref{Lagr}) 
becomes valid). In any case this makes the constant $C$ in 
Eq. (\ref{osc}) very small, practically zero at present and greatly 
reduces possible phase space for this DE model.

Still, if this problem is solved and $C$ may be put zero somehow, 
then a viable FRW background is obtained once the condition
(\ref{ineq2}) is satisfied. Moreover, since it follows 
from the present observational data that $\Lambda(R_1)$ is not too 
much different from $\Lambda(\infty)$, one can choose $x_1$ close to 
$2\lambda$. Then the whole FRW evolution is analytically described by 
the first iteration formula (\ref{ind}) where the background FRW 
metric is that of the $\Lambda$CDM model:
\begin{equation}
a^{(0)}\propto \sinh^{2/3}\left(\frac{3}{2}H_0t\right)~,~~
H^{(0)}=H_0\coth\left(\frac{3}{2}H_0t\right)~,~~R^{(0)}=3H_0^2
\left(4+\frac{1}{\sinh^2(\frac{3}{2}H_0t)}\right)
\label{lcdm}
\end{equation}
where $H\equiv \dot a/a,~H_0^2=\Lambda(\infty)/3=\lambda R_0/6$. 
Knowing $\delta R_{ind}$, it is straightforward to obtain $\delta H$
and $\delta a/a$.     

\section{Laboratory and Solar System tests and dynamics of 
inhomogeneities}

The main problem with laboratory and Solar system tests of gravity
for all $f(R)$ DE models irrespective of the form of $f(R)$ is that
this class of models represents the limiting case $\omega\to 0$ of
scalar-tensor gravity as was mentioned above. As a result, would 
scalaron be massless, an additional `fifth force' would show itself 
in laboratory experiments and the first post-Newtonian parameters
would have the values $\beta=1,~\gamma=1/2$ that is not admissible.
However, this problem is qualitatively the same as it occurs for the 
string theory dilaton which in the massless low-energy limit
corresponds to  
scalar-tensor gravity with $\omega=-1$ that is excluded, too. As is
well known, that problem is solved by assuming that the dilaton is 
sufficiently massive. The same one has to assume about the scalaron 
in $f(R)$ DE models. Namely, let us choose model parameters in such 
a way that for any laboratory or Solar system test of gravity having 
a characteristic scale $L$ and made in an environment with some 
non-relativistic matter density $\rho_m$, the scalaron mass 
satisfies the strong inequality $M(R(\rho_m))L\gg 1$. Under this
condition deviations from the equation $R=8\pi GT_m$ (and the Poisson
equation, too) are small both inside and outside the Sun and other
compact bodies. There have been hopes that this requirement may be 
circumvented using the so called ``Chameleon'' effect \cite{KW04} 
(see also \cite{MB04}). However, until recently it 
has not been shown to what extent can it be important in construction 
of viable $f(R)$ gravity models, see e.g. \cite{FTBM06}. For the 
model (\ref{f}), $M(\rho_m)\propto \rho_m^{n+1}$ for $R\gg R_0$ 
with $M(R_0)\sim R_0^{1/2}\sim 10^{-28}$ cm$^{-1}$.

 For the most recent and best Cavendish-type experiment \cite{KCA07},
taking $L\approx 50\, \mu$m and $\rho_m\approx 10^{-12}$ g cm$^{-3}$
(corresponding to a vacuum of $\approx 10^{-6}$ torr achieved there),
we obtain the sufficient condition $n\ge 1$ (rounded to a larger 
integer). In the case of light deflection or the Shapiro time delay 
by the Sun, the main contribution to $\gamma -1$ is from distances 
$r\sim R_{\odot}\approx 7\times 10^{10}$ cm. If we take the Solar 
corona density at this distance ($\sim 10^{-15}$ g cm$^{-3}$) 
as $\rho_m$, then already $n\ge 0.5$ would be sufficient. In the case 
$0.5<n< 2$
the condition $M(R)\,r\gg 1$ can be violated farther from the Sun. 
However, the scalar component of Solar gravitational field has been 
already screened by the Yukawa damping factor $\exp\left(-\int M(r)\,
dr\right)$ by that 
distance. Clearly more careful calculation is needed here with the
detailed account of interplanetary matter profile in the Solar
system. The same refers to such tests like Lunar laser ranging.
In any case, all known laboratory and Solar system tests of gravity
are certainly satisfied for $n\ge 2$ and probably this condition may be
softened up to $n\ge 1$. However, it will be shown below that there 
is no necessity in such softening due to limits from the large-scale 
structure of the Universe.

Finally, let us turn to the evolution of weak inhomogeneities in the 
linear regime. It follows from equations for perturbations, either
obtained for scalar-tensor gravity with the limit $\omega \to 0$ taken
in them, see e.g. \cite{BEPS00}, or from those directly derived for 
$f(R)$ theory (\cite{KMP87,HN01,SHS07} and other papers), that the 
equation for density perturbations $\delta_m\equiv \delta\rho_m/
\rho_m$ in the non-relativistic matter component (cold dark matter 
+ baryons) during the matter-dominated stage has the form
\begin{equation}
\ddot \delta_m+2H\dot\delta_m-4\pi G_{eff}\rho_m\delta_m =0
\label{inhom}
\end{equation}
in the limit $k\gg aH$ where $k\equiv |{\bf k}|$ (spatial dependence
$\exp(i{\bf kr})$ is assumed) and $G_{eff}=G/f'(R)$ for $k\ll M(R)a$ 
and $G_{eff}=4G/3f'(R)$ for $k\gg M(R)a$ ($f'(R)\approx 1$ for 
$R\gg R_0$). This increase of $G_{eff}$ in $4/3$ times is just how
the 'fifth force' due to additional scalar gravity shows itself at 
large scales and low matter density. 

Because of this, $\delta_m$ grows as usually, $\propto t^{2/3}$,
before the time moment $t_k$ when $k=M(R)a$, but after that this law 
changes
to $\delta_m\propto t^{\frac{\sqrt{33}-1}{6}}$ that continues up
to $t_{\Lambda}$ -- the end of the matter-dominated stage when 
$\ddot a=0$. Using (\ref{mat-back}), we obtain $t_k\propto
k^{-1/2(n+\frac{2}{3})}$. As a result, $\delta_m(k)$ acquires 
an additional growth factor during the matter-dominated stage
(compared to the $\Lambda$CDM model) proportional to
\begin{equation}
\left(\frac{t_{\Lambda}}{t_k}\right)^{\frac{\sqrt{33}-1}{6}-
\frac{2}{3}}\propto k^{\frac{\sqrt{33}-5}{4\,(3n+2)}}~.
\label{factor}
\end{equation}
This additional increase occurs at small redshifts and is not seen
in CMB fluctuations (apart from some features in a few low multipoles 
which are bounded by cosmic variance). E.g., for $n=2,~k/a_{now}
H_{now}=300$ and $z_{\Lambda}=0.7$, the redshift $z_k(t_k)=2.5$. As a 
result, there arises some discrepancy
between values of the slope $n_s$ of the primordial power spectrum 
determined from galaxy surveys on one side (assuming the standard 
evolution of perturbations) and CMB fluctuations on the other:
\begin{equation}
\Delta n_s = n_s^{(gal)}-n_s^{(CMB)}= \frac{\sqrt{33}-5}{2\,(3n+2)}~.
\label{delta}
\end{equation}
For comparison, $\Delta n_s$ is equal to 0.074 for $n=1$ and 0.047
for $n=2$. Note that the limit $n\to 0$ corresponds not to the 
$\Lambda$CDM model but to $F(R)\propto \ln R$ at large $R$. 

At present no such discrepancy is seen, see e.g. \cite{T06}, so 
we may conservatively bound $\Delta n_s<0.05$ that leads to $n\ge 2$
in (\ref{f}). Of course, a more exact numerical calculation is needed
for larger values of $n$ since then the values of $z_k$ lie rather close 
to unity, so the formula (\ref{delta}) becomes too approximate.

\section{Conclusions and discussion}

Thus, in contrast to numerous unsuccessful previous attempts to 
construct a viable DE model in $f(R)$ gravity using a function $f$ 
divergent or non-analytic at $R=0$, it appears possible to achieve
this goal
with a regular $f(R)$ satisfying the condition $f(0)=0$ which means the
absence of a 'bare' cosmological constant in flat space-time. On the
other hand, DE in this model behaves itself as an effective
cosmological constant for large $R$ if $F(R)=f-R\to const$ at $R\to
\infty$. This model passes laboratory and Solar system tests of
gravity if its parameter $n$ is sufficiently large ($n\ge 2$ seems to 
be the sufficient, but probably not necessary condition), though
analysis of gravitational radiation from double pulsars may add some
new restriction. But it is clear already that limits from large-scale
structure arising due to the anomalous growth of linear perturbations 
at recent redshifts are more critical and lead to a stronger limit on 
$n$. Just the opposite, any discrepancy $\Delta n_s$ between values
of the slope of the primordial perturbation spectrum obtained from
galaxy and CMB data may serve as a strong argument for such model. 
Then Eq. (\ref{delta}) directly relates $\Delta n_s$ to $n$.

However, more deep theoretical analysis of this class of models
has uncovered its new serious problem not considered before: how to
avoid an overabundance of new scalar particles arising in $f(R)$
gravity (dubbed scalarons) which can be generated in the very
early Universe. Mathematically this means that the coefficient $C$
in Eq. (\ref{osc}) should be practically zero at present, i.e.
almost the whole degree of freedom has to be suppressed.
Otherwise, a FRW solution in this model cannot have sufficiently
long radiation- and matter-dominated stages because it
hits the weak singularity where $f''(R)=0$. Note that this difficulty 
is of a more subtle type than those of linear stability considered
before. It is rather a problem of a measure of initial conditions
in the early Universe leading to the standard cosmological evolution 
almost up to the present time. Anyway it remains an important topic 
for further study. 
  
While this paper was being prepared for publication, two papers 
\cite{HS05,AB05} appeared where similar DE models in $f(R)$ gravity 
possessing the property $f(0)=0$ were proposed. Our model is closer to 
that in \cite{HS05}. 

\acknowledgments

The research was partially supported by the Russian Foundation for
Basic Research, grant 05-02-17450, by the Research Programme
``Astronomy'' of the Russian Academy of Sciences and by the scientific 
school grant 1157.2006.2. The author thanks the Yukawa Institute for 
Theoretical Physics, Kyoto University for hospitality during the
period when this project was finished.

\end{document}